\documentclass[12pt]{article}

\usepackage{graphicx}

\begin{document}

\title{Interaction of a soliton with a local defect in a fiber Bragg grating}

\author{William C. K. Mak$^{1}$, Boris A. Malomed$^2$, and Pak L. Chu$^{1}$\\
$^1$Optoelectronic Research Centre, Department of Electronic
Engineering, City University of Hong Kong, Tat Chee Avenue,
Kowloon, Hong Kong\\
$^2$Department of Interdisciplinary Studies,
Faculty of Engineering, Tel Aviv University,
Tel Aviv, Israel \\}

\begin{center}
Abstract
\end{center}
We study the interactions of a Bragg-grating soliton with a localized defect
which is a combined perturbation of the grating and refractive index. A
family of exact analytical solutions for solitons trapped by the delta-like
defect is found. Direct simulations demonstrate that, up to the numerical
accuracy available, the trapped soliton is stable at a \emph{single} value
of its intrinsic parameter $\theta $. Depending on parameter values,
simulations of collisions between moving solitons and the defect show that
the soliton can get captured, pass through, or even \emph{bounce from} the
defect. If the defect is strong and the soliton is heavy enough, it may
split into three fragments: trapped, transmitted, and reflected ones.

\maketitle

\newpage

\section{Introduction}

Optical fibers carrying a Bragg grating (BG) written on their cores have
attracted a great deal of attention for various reasons \cite{Kashyap}.
Fibers equipped with homogeneous or chirped BG are a basis for the design of
many elements used in optical telecommunications, such as filters \cite
{filter}, add-drop WDM multiplexers \cite{multiplexer1,multiplexer2},
dispersion compensators \cite{compensator}, and others. They also find
important applications as sensors, since the optical properties of the
fiber-based BG are strongly affected by local stress, see, e.g., a recent
work \cite{recent} and reviews \cite{sensor1,sensor2}. Besides that,
BG-carrying fibers are interesting in their own right, as a unique physical
medium supporting various optical excitations. In particular, strong
dispersion induced by BG opens a way to observe slow light in this medium,
which is a topic of great current interest \cite{slow1,slow2}.

The BG-induced dispersion in combination with the Kerr nonlinearity of the
fiber give rise to localized pulses, which are usually called gap solitons
(or, more generically, BG solitons \cite{Sterke}). After being predicted
analytically \cite{Aceves,Demetri} (see also the review \cite{Sterke}), BG
solitons have been observed in experiments \cite{experiment1,experiment2}.
They have attracted a lot of interest as their dynamical properties are very
different, in many respects, from the properties of usual solitons in
optical fibers (except for the regime in which the amplitude of the BG
soliton is small, then it is virtually equivalent to a usual
nonlinear-Schr\"{o}dinger soliton \cite{Litchinitser}). In particular, quest
for very slow solitons, that are expected to exist in fiber BGs \cite
{Demetri}, is a challenge \cite{slowsoliton}.

The subject of this work is interaction of a soliton with a local defect in
the BG-carrying fiber. A local defect in this medium may trap light via
four-wave mixing without the formation of a soliton per se \cite{trapping},
but the interaction of solitons with defects is especially interesting \cite
{Weinstein}. First, it is a fundamental physical problem -- in particular,
because the defect may help to stop the soliton and thus to create a
self-sustained pulse of standing light (leaving aside the problem of the
fiber-loss compensation, which is a subject of another work \cite{PREnew}).
Second, narrow soliton pulses interacting with local inhomogeneities may
increase the sensitivity when the grating is used as a sensor. Moreover, the
trapping of soliton in a fiber BG may be used to design optical-memory
devices. Further potential applications include optical delay lines and
oscillators, see the concluding section of this paper.

The objective of the present work is a detailed study of the interaction of
BG solitons with a local point-like defect. In particular, it will be shown
that a moving soliton can be captured by the defect. We will also study the
stability of trapped solitons, concluding that they may be both unstable and
stable, depending on the values of the system's parameters. It is necessary
to mention that the interaction between a BG soliton and local defect of the
grating was considered in a very recent paper \cite{Weinstein}, which was
chiefly focused on a detailed analysis of the case of finite-size defects.
We, instead, consider point-like defects, which makes it possible to find an 
\emph{exact} solution for trapped solitons, and an analytical potential of
the interaction between a moving soliton and the defect. The study of the
soliton-defect interaction reveals features that were not reported in Ref. 
\cite{Weinstein} -- most notably, reversal of the sign of the interaction
from attraction to repulsion in some cases. Another interesting property
which we present here is the fact that a stable stationary trapped soliton
is, as a matter of fact, an \textit{attractor}: up to the accuracy of
numerical simulations, it is found to exists at a single value of its
energy, but initial solitons from a broad area in the system's parameter
space relax to this stable state.

The paper is organized as follows: in section 2, the model is formulated,
and analytical results are presented in section 3. Section 4 contains
numerical results for the stability of trapped solitons, which show the
existence of the attractor. In section 5, simulations of collisions of
moving solitons with the defect are presented, the most remarkable feature
being the above-mentioned reversal of the sign of the interaction; this, in
particular, gives rise to two different regions in the parameter space,
where the moving soliton is captured or repelled by the defect. The paper is
concluded by section 6.

\section{The model}

We approximate a local disruption (or enhancement) of the resonant Bragg
grating due to the presence of a defect by a $\delta $-function perturbation
added to the standard nonlinear BG model for the right- and left-traveling
waves $u(x,t)$ and $v(x,t)$. Taking into regard that the local perturbation
may also affect the fiber's refractive index, cf. Ref. \cite{Weinstein}, we
arrive at the following model, written in the usual notation \cite
{Sterke,Aceves,Demetri}:

\begin{eqnarray}
i\frac{\partial u}{\partial t}+i\frac{\partial u}{\partial x}+v+(\frac{1}{2}
|u|^{2}+|v|^{2})u &=&\kappa \delta (x)v+\lambda \delta (x)u,  \label{utux} \\
i\frac{\partial v}{\partial t}-i\frac{\partial v}{\partial x}+u+(\frac{1}{2}
|v|^{2}+|u|^{2})v &=&\kappa \delta (x)u+\lambda \delta (x)v,  \label{vtvx}
\end{eqnarray}
where the ordinary ratio between the self-phase-modulation (SPM) and
cross-phase-modulation (XPM) coefficients, $1:2$, is adopted, the local
defect is placed at the point $x=0$, and the real constants $\kappa >0$ and 
$\lambda $ measure, respectively, the local suppression of the grating 
($\kappa <0$ corresponds to a locally enhanced Bragg reflection) and local
perturbation of the refractive index.

It should be mentioned that the region with a locally changed refractive
index may also give rise to partial reflection, independently from that
induced by BG. The corresponding perturbation is described by the
cross-coupling linear terms proportional to the $\delta $-function, which
are similar to those (proportional to $\kappa $) that are already present in
Eqs. (\ref{utux}) and (\ref{vtvx}); however, the coefficients in front of
these additional terms may be complex, due to the phase shift of the
reflected wave. Thus, in the most general case, the terms $\kappa \delta (x)v
$ and $\kappa \delta (x)u$ with real $\kappa $ in Eqs. (\ref{utux}) and (\ref
{vtvx}) can be replaced, respectively, by $\left( \kappa _{1}+i\kappa
_{2}\right) \delta (x)v$ and $\left( \kappa _{1}-i\kappa _{2}\right) \delta
(x)u$, with two real parameters, $\kappa _{1}$ and $\kappa _{2}$. However,
in this work we confine the consideration to the case of real $\kappa $. As
for the parameter $\lambda $, it must be strictly real in the conservative
model (its imaginary part accounts for a local gain, if it is present in the
model \cite{PREnew}).

If $\kappa $ is positive and small, the corresponding perturbation terms in
Eqs. (\ref{utux}) and (\ref{vtvx}) imply that the Bragg reflection is
disrupted in a region whose size is much smaller than a characteristic
length necessary for the complete reflection, which is, typically, $\sim \,1$
mm \cite{Sterke,experiment1,experiment2}, and which also determines a
characteristic intrinsic spatial scale of the soliton. If, on the other
hand, $\kappa \sim 1$, this implies that the size of the disruption region
is comparable to the reflection length. In the latter case, an exact form of
the spatially distributed perturbation should be, strictly speaking, taken
into regard (see Ref. \cite{Weinstein}). Nevertheless, the use of the 
$\delta $-function approximation may be justified in this case too, as the
fundamental features of the interaction of a soliton with the inhomogeneity
may be weakly sensitive to the exact shape of the inhomogeneity. The latter
property is well known to be true in many other applications of the soliton
perturbation theory \cite{KM}. In accordance with this, in the numerical
part of the work we will assume that $\kappa $ takes values between $0$ and 
$1$. On the other hand, in the case $\kappa <0$, $\left| \kappa \right| $ may
take any value, as in this case the $\delta $-function terms in Eqs. (\ref
{utux}) and (\ref{vtvx}) have the literal meaning of a local perturbation in
the form of the enhanced Bragg reflection.

It is relevant to mention that this difference in the interpretation of
positive and negative values of $\kappa $ is very similar to the well-known
difference between the positive and negative values of a perturbation
parameter that determines the local suppression or enhancement
(corresponding, respectively, to the so-called micro-resistor or
micro-short) of the tunneling supercurrent in the standard model of a long
Josephson junction, which is based on the sine-Gordon equation with a 
$\delta $-functional perturbation \cite{KM}. Indeed, a micro-resistor may
only make the existing critical density of the supercurrent smaller, while a
micro-short may arbitrarily increase the local value of the critical
density. As concerns the refractive-index perturbation parameter $\lambda $
in Eqs. (\ref{utux}) and (\ref{vtvx}), it may take arbitrary values, as it
literally accounts for a local perturbation.

Notice that Eqs. (\ref{utux}) and (\ref{vtvx}) can be derived from the
Hamiltonian, 
\[
H=\int_{-\infty }^{+\infty }\left[ \frac{1}{2}i\left( -u^{\ast }\frac{
\partial u}{\partial x}+v^{\ast }\frac{\partial v}{\partial x}+u\frac{
\partial u^{\ast }}{\partial x}-v\frac{\partial v^{\ast }}{\partial x}
\right) -\frac{1}{4}\left( \left| u\right| ^{4}+\left| v\right| ^{4}\right)
-\left| u\right| ^{2}\left| v\right| ^{2}-\left( u^{\ast }v+uv^{\ast
}\right) \right] +H_{\mathrm{int}}\,. 
\]
In this expression, the asterisk stands for the complex conjugation, the
overall sign of $H$ is chosen so that the SPM\ and XPM terms correspond to
self-focusing, and the perturbation part of the Hamiltonian, which accounts
for the interaction of the soliton with the local defect, is 
\begin{equation}
H_{\mathrm{int}}=\left[ \kappa \left( u^{\ast }v+uv^{\ast }\right) +\lambda
\left( \left| u\right| ^{2}+\left| v\right| ^{2}\right) \right] |_{x=0}\,.
\label{int}
\end{equation}

\section{Analytical results}

We start by seeking for stationary solutions to Eqs.~(\ref{utux}) and (\ref
{vtvx}) in the form 
\begin{equation}
u=U(x)\,\exp \left( -i\omega t\right) ,v=V(x)\,\exp \left( -i\omega t\right)
,  \label{stationary}
\end{equation}
which leads to equations

\begin{eqnarray}
\omega U+iU^{\prime }+V+(\sigma |U|^{2}+|V|^{2})U &=&\kappa \delta
(x)V+\lambda \delta (x)U,  \label{sutux} \\
\omega V-iV^{\prime }+U+(\sigma |V|^{2}+|U|^{2})V &=&\kappa \delta
(x)U\,+\lambda \delta (x)V,  \label{svtvx}
\end{eqnarray}
the prime standing for $d/dx$. A family of exact solutions to the
unperturbed equations (\ref{sutux}) and (\ref{svtvx}), with $\kappa =\lambda
=0$, is well known \cite{Aceves,Demetri} (these solutions are frequently
called gap solitons): 
\begin{equation}
U(x)=\sqrt{\frac{2}{3}}(\sin \mathrm{\,}\theta )\,\mathrm{sech}\left( \left(
x-\xi \right) \cdot \sin \mathrm{\,}\theta -\frac{i}{2}\theta \right)
,\,V=-U^{\ast },  \label{exact}
\end{equation}
where $\xi $ is the coordinate of the soliton's center, and $\theta $ is a
real parameter taking values 
\begin{equation}
0<\theta <\pi ,  \label{interval}
\end{equation}
which also determines the frequency in Eqs. (\ref{stationary}), 
\begin{equation}
\omega =\cos \,\theta .  \label{omega}
\end{equation}

\subsection{An exact solution for the pinned soliton}

An \emph{exact} solution for a stationary soliton trapped by the local
defect can be found in the form [cf. Eq. (\ref{exact})] 
\begin{equation}
U=\sqrt{\frac{2}{3}}(\sin \theta )\,\mathrm{sech}\left[ \left( x+a\,\mathrm{
\ sgn\,}x\right) \cdot \sin \mathrm{\,}\theta -\frac{i}{2}\theta \right]
,\,V=-U^{\ast },  \label{pinned}
\end{equation}
where an expression for the parameter $a$ is obtained by integrating Eqs. 
(\ref{sutux}) and (\ref{svtvx}) in an infinitesimal vicinity of the point 
$x=0 $: 
\begin{equation}
\tanh \left( a\sin \theta \right) \,=\sqrt{\frac{\kappa -\lambda }{\kappa
+\lambda }}\frac{\tanh \left( \sqrt{\kappa ^{2}-\lambda ^{2}}/2\right) }{
\tan \left( \theta /2\right) }.  \label{a}
\end{equation}

\noindent An example of this exact solution, with $\kappa=0.4$ and 
$\theta=0.5\pi$, is shown in Fig.~1.

In Eq. (\ref{a}), we assume that $\kappa ^{2}>\lambda ^{2}$ (the opposite
case will be considered separately below), and $\sqrt{\kappa ^{2}-\lambda
^{2}}$ is realized in such a way that it has the same sign as $\kappa$
itself. In particular, in the case $\lambda =0$, Eq. (\ref{a}) takes the
form 
\begin{equation}
\tanh \left( a\sin \theta \right) =\frac{\tanh \left( \kappa /2\right)} 
{\tan \left( \theta /2\right) }\,.  \label{a1}
\end{equation}

The intensity of the field at the central point of the trapped soliton is 
\begin{equation}
\left| U\left( x=0\right) \right| ^{2}=\frac{2}{3}\frac{\sin ^{2}\theta }
{\sinh ^{2}\left( a\sin ^{2}\theta \right) +\cos ^{2}\theta }\,.
\label{centerpoint}
\end{equation}
Note that $\left| U(x)\right| $ is continuous in the solution (\ref{pinned})
across the point $x=0$, while the phase of $U(x)$ has a discontinuity at
this point.

If $\kappa <0$, Eqs. (\ref{a}) and (\ref{a1}) yield $a<0$, hence, as it
follows from Eq. (\ref{pinned}), in this case $\left| U(x)\right| $ has two
symmetric local maxima at $x=\mp \,a$, and a local minimum at the central
point $x=0$. On the contrary to this, if $\kappa >0$, we have $a>0$, which
implies that the solution has a single maximum at $x=0$. These features can
be readily understood by dint of the interaction Hamiltonian (\ref{int}).
Indeed, together with the relation $V=-U^{\ast }$, see Eq. (\ref{exact}),
the expression (\ref{int}) yields the following value of the interaction
Hamiltonian for the trapped soliton: 
\begin{equation}
\left( H_{\mathrm{int}}\right) _{\mathrm{trapped}}=-2\left( \kappa -\lambda
\right) \left| U\left( x=0\right) \right| ^{2}.  \label{Hint_pin}
\end{equation}
With regard to the constraint $\kappa ^{2}>\lambda ^{2}$ adopted above, we
conclude that this contribution increases the net Hamiltonian if $\kappa $
is negative, and decreases it in the opposite case. Therefore, following the
general principle stating that the ground state of a system corresponds to
the minimum of $H$, the system tries to make $\left| U\left( x=0\right)
\right| ^{2}$ smaller if $\kappa <0$, and larger if $\kappa >0$.

The result for the case $\kappa ^{2}<\lambda ^{2}$ can be obtained from Eq. 
(\ref{a}) by analytical continuation [or directly from Eqs. (\ref{sutux}) and
(\ref{svtvx})]: 
\begin{equation}
\tanh \left( a\sin \theta \right) =-\sqrt{\frac{\lambda -\kappa }{\lambda
+\kappa }}\frac{\tan \left( \sqrt{\lambda ^{2}-\kappa ^{2}}/2\right) }{\tan
\left( \theta /2\right) }\,,  \label{a2}
\end{equation}
where $\mathrm{sgn}\left( \sqrt{\lambda ^{2}-\kappa ^{2}}\right) \equiv 
\mathrm{sgn}\,\lambda $. In the particular case $\kappa =0$, Eq. (\ref{a2})
simplifies to 
\begin{equation}
\tanh \left( a\sin \theta \right) =-\frac{\tan \left( \lambda /2\right)}
{\tan \left( \theta /2\right) },  \label{a3}
\end{equation}
cf. Eq. (\ref{a1}). Equations (\ref{a2}) and (\ref{a3}) yield $a>0$, i.e.,
the single maximum of $\left| U(x)\right| $ at $x=0$, if $\lambda $ is
negative, and $a<0$, i.e., a local minimum at $x=0$ and two maxima at $x=\mp
a$, if $\lambda $ is positive. Taking into regard the present constraint 
$\kappa ^{2}<\lambda ^{2}$, it is easy to check that these conclusions are
again in accordance with the principle of minimization of the interaction
Hamiltonian (\ref{Hint_pin}).

In the special case $\kappa =-\lambda $, both Eqs. (\ref{a}) and (\ref{a2})
amount to 
\begin{equation}
\tanh \left( a\sin \theta \right) =\frac{\kappa }{\tan \left( \theta
/2\right) }\,\equiv -\frac{\lambda }{\tan \left( \theta /2\right) }\,.
\label{a0}
\end{equation}
Lastly, in the case $\kappa =\lambda $, both expressions (\ref{a}) and (\ref
{a2}) yield $a=0$, which is easy to understand: due to the relation 
$V=-U^{\ast }$, the unperturbed soliton (\ref{exact}) remains an exact
solution to Eqs. (\ref{sutux}) and (\ref{svtvx}) if $\kappa =\lambda$.

Obviously, the expressions (\ref{a}), (\ref{a1}), and (\ref{a2}) through 
(\ref{a0}) make sense if they yield $\left| \tanh \left( a\sin \theta \right)
\right| <1$, otherwise the solution for the trapped soliton does not exist.
The latter restriction means that the exact solution exists not in the whole
interval (\ref{interval}) in which the unperturbed solution (\ref{exact})
was found, but rather in a narrower band, 
\begin{equation}
\theta _{\min }<\theta <\pi ,  \label{narrow}
\end{equation}
where, in both cases $\kappa ^{2}\,^{<}_{>}\, \lambda ^{2}$, 
\begin{equation}
\theta _{\min }=2\tan ^{-1}\left( \left| \sqrt{\frac{\kappa -\lambda} 
{\kappa +\lambda }}\tanh \left( \sqrt{\kappa ^{2}-\lambda ^{2}}/2\right)
\right| \right) .  \label{min}
\end{equation}
This expression takes the form $\theta _{\min }=\lambda $ in the case 
$\kappa =0$, and in the case $\lambda =-\kappa $ it is $\theta _{\min }=2\tan
^{-1}\left( \left| \kappa \right| \right) $.

The existence limit for the trapped solitons, as given by Eqs. (\ref{narrow})
and (\ref{min}), was also verified by numerical methods, solving Eqs. (\ref
{sutux}) and (\ref{svtvx}) with some approximations for $\delta (x)$, see
details below. The numerical solutions (not shown here, as they are not
especially interesting by themselves) demonstrate good agreement with the
analytically predicted existence limit.

\subsection{A remark about the soliton stability}

Some (incomplete) conclusions about the stability of the trapped solitons
can be made on the basis of the known Vakhitov-Kolokolov (VK) criterion \cite
{VK}, which states that a necessary condition of the stability of the
soliton is $dE/d\omega <0$, where $\omega $ is the soliton's frequency, see
Eq. (\ref{omega}), and $E\equiv \int_{-\infty }^{+\infty }\left[ \left|
u(x)\right| ^{2}+\left| v(x)\right| ^{2}\right] dx$ is the energy (norm) of
the solution. In the general case, the expression for the energy of the
exact solution (\ref{pinned}) is cumbersome, but it takes a simple form in
the case $\lambda =0$, on which the numerical simulations will be focused
below: 
\begin{equation}
E=\frac{8}{3}\left( \theta -\left[ \frac{\pi }{2}-\sin ^{-1}\left( \mathrm{\
sech\,}\kappa \right) \right] \mathrm{sgn}\,\kappa \right) .  \label{E}
\end{equation}

The calculation of the derivative $dE/d\omega $ by means of Eqs. (\ref{E})
and (\ref{omega}) shows that all the trapped-soliton solutions satisfy the
VK criterion. However, this result is not sufficient for claiming stability
of the solutions. Indeed, in the unperturbed case ($\kappa =0$) the VK
criterion per se predicts that all the solitons (\ref{exact}) are stable.
However, studies of the stability, performed originally within the framework
of the variational approximation \cite{Rich}, and then by means of accurate
numerical methods \cite{Barash}, had demonstrated that, in fact, the
interval (\ref{interval}) in which the solitons exist is divided into two
parts: they are stable if 
\begin{equation}
0<\theta <\theta _{\mathrm{cr}},\,\theta _{\mathrm{cr}}\approx 1.011\cdot
\left( \pi /2\right) ,  \label{stability}
\end{equation}
and unstable if $\theta _{\mathrm{cr}}<\theta <\pi $. Notice that $\theta_{
\mathrm{cr}}$ is just slightly larger than $\pi /2$, which is the value at
which the gap soliton (\ref{exact}) has the largest amplitude and smallest
width. This example clearly suggests that direct investigation of the
solitons' stability is necessary in the present model too, which will be
done by means of numerical methods in section 4.

\subsection{The potential of the soliton-defect interaction}

Besides the exact findings presented above, more general analytical results
can be obtained in an approximate form, assuming that $\kappa $ and $\lambda 
$ are small parameters. In the lowest (adiabatic) approximation of the
perturbation theory \cite{KM}, one can find an effective potential of the
interaction between the soliton, treated as a quasi-particle, and the local
defect. To this end, the unperturbed wave form, based on Eqs. (\ref
{stationary}) and (\ref{exact}), is substituted into the expression (\ref
{Hint_pin}) for the interaction Hamiltonian, which yields 
\begin{equation}
U_{\mathrm{int}}(\xi )=\frac{8}{3}\frac{\left( \lambda -\kappa \right) \sin
^{2}\theta }{\cosh \left( 2\xi \sin \theta \right) +\cos \theta }\,
\label{U}
\end{equation}
(recall $\xi $ is the coordinate of the soliton's center). In particular, in
the case $\lambda =0$ that will be considered below, the potential (\ref{U})
clearly implies attraction in the case $\kappa >0$, and repulsion in the
opposite case. The latter circumstance strongly suggests that the trapped
soliton cannot be stable in the case $\kappa <0$, $\kappa ^{2}>\lambda ^{2}$
(nor in the case $\lambda >0$, $\lambda ^{2}>\kappa ^{2}$).

\section{Numerical simulations of the stability of the pinned soliton}

\subsection{The approximation for the delta-function}

To simulate the stability of the trapped solitons and the interaction of a
free soliton with the local defect, we have to adopt an approximation to
represent the $\delta $-functions in Eqs.~(\ref{utux}),(\ref{vtvx}), (\ref
{sutux}) and (\ref{svtvx}). In this work, we use a numerical scheme in which
the coordinate $x$ is represented by $501$ grid points $x_{j}$, 
$j=-250,...,-1,0,+1,...+250$. As an approximation to the $\delta $-function,
we have defined the following function on a set of $2N+1$ grid points in the
central part of the integration domain, located symmetrically around zero:

\begin{equation}
\widetilde{\delta }\left( x_{n-\left( N+1\right) }\right) \equiv 
\begin{array}{ll}
A\,\mathrm{cos}\,\left( \frac{n-\left( N+1\right) }{2N+1}\pi \right) & 
\mathrm{for}\;n=1,\ldots ,2N+1, \\ 
0 & \mathrm{otherwise.}
\end{array}
\label{approxd}
\end{equation}
\noindent The normalization factor is defined so as to maintain the
canonical normalization of the $\delta $-function, $\int_{-\infty }^{+\infty
}\widetilde{\delta }(x)dx\equiv \sum_{j}\widetilde{\delta }\left(
x_{j}\right) \Delta x=1$, which yields 
\begin{equation}
A=\left[ \Delta x\sum_{n=1}^{2N+1}\mathrm{cos}\,\left( \frac{n-\left(
N+1\right) }{2N+1}\pi \right) \right] ^{-1},  \label{A}
\end{equation}
$\Delta x$ being the spacing of the grid (in fact, it was $0.04)$.

In most cases presented below, we use $N=2$ [then Eq. (\ref{A}) with $\Delta
x=0.04$ yields $A=\left( \left( \allowbreak 1+\sqrt{5}\right) \Delta
x\right) ^{-1}\approx \allowbreak 7.\,\allowbreak 726$], which makes the 
$\delta $-function quite narrow indeed. Besides that, we also used another
approximation for the $\delta $-function, in which a simple rectangular
regularization was used instead of (\ref{approxd}). In some cases, sharp
edges of the rectangular approximation produced problems with the
convergence, but in those cases when results converged, they were virtually
the same as generated by the approximation (\ref{approxd}).

It should be noted that, in the case $\kappa A>1$ [recall $\kappa $ is the
perturbation parameter in Eqs. (\ref{utux}) and (\ref{vtvx})], the local
value of the coefficient $\kappa \widetilde{\delta }(x)$ may exceed $1$,
which is, strictly speaking, an unphysical situation, as it was discussed
above. However, we consider such a situation as modeling a physical one with
a broader defect and smaller amplitude, when the integral strength of the
perturbation, $\kappa \int_{-\infty }^{+\infty }\widetilde{\delta }(x)dx$,
takes the same value, while the coefficient $\kappa \widetilde{\delta }(x)$
does not exceed $1$. Comparing results produced in simulations by the
formally unphysical perturbation, and by its physical counterpart, we have
verified that this assumption is, to a large extent, correct, with some
minor differences observed only when the defect strength $\kappa $ is small.
This issue will be further discussed below.

\subsection{The stationary soliton}

In simulations of the stability of the trapped solitons and the interaction
of a free soliton with the local defect (the interaction is the subject of
the next section), we concentrated on the case $\lambda =0$, which is the
most interesting one, as it focuses on the effects induced by the defect in
the BG proper. We will mainly assume $\kappa >0$ \ (although the opposite
case will also be briefly considered), which corresponds to the most natural
situation with the local perturbation suppressing the local Bragg
reflectivity (rather than enhancing it). This is also justified because, as
conjectured above and shown below, all the trapped solitons are unstable in
the case $\kappa <0$.

Our first objective is to analyze the stability of a stationary soliton
trapped by the local defect. To this end, instead of using the exact
solution given by Eqs. (\ref{pinned}) and (\ref{a}) through (\ref{a0}), we
directly constructed trapped-soliton solutions to Eqs. (\ref{sutux}) and 
(\ref{svtvx}), with the regularized $\delta $-function taken as per Eq.~(\ref
{approxd}), by means of the Newton-Raphson method. An example of the thus
found stationary soliton is shown in Fig. 1, along with the exact solution 
(\ref{pinned}), for $\kappa =0.4$ and $\theta =\pi /2$. After obtaining the
stationary solutions in the numerical form, it was found that they are
always close to the analytical solution for the ideal $\delta $-function
given by Eqs. (\ref{pinned}) and (\ref{a}), so that the value of the
parameter $\theta $ corresponding to each numerically found solution could
be easily identified.

\subsection{Stability simulations}

The stability of the stationary solutions was directly simulated by means of
the split-step method applied to Eqs.~(\ref{utux}) and (\ref{vtvx}),
employing the fast Fourier transform. First of all, in the case $\kappa <0$
the numerical solution confirms that, as it follows from Eq. (\ref{pinned}), 
$\left| U(x)\right| $ has two symmetric local maxima at $x=\mp \,a$, and a
local minimum at the central point $x=0$. Dynamical simulations confirm that 
\emph{all} the trapped solitons with $\kappa <0$ are unstable, as it was
assumed above on the basis of the perturbation theory. The simulations
further show that, as a result of the instability, two local maxima at 
$x=\mp \,a$ start to separate and eventually move away from the defect, so
that the trapped soliton splits into two free separating ones. Since all the
trapped solitons are unstable for $\kappa <0$, the investigation of this
case was not carried out further.

Typical results illustrating the stability of the trapped solitons in the
case $\kappa >0$ are displayed in Fig.~2, which shows the evolution of the
initial pulse in the form of the trapped soliton (\ref{pinned}), with its
center placed at $x=0$, the value of the perturbation parameter being 
$\kappa =0.08$ (note that this case is a ``strictly physical'' one, as it
corresponds to $\kappa A<1$, see above). In the series of panels displayed
in Fig. 2, $\theta $ is given values $\theta _{\mathrm{in}}=0.4\pi ,\,0.5\pi
,\,0.7\pi ,0.9\pi $. The left part of each panel shows a ``side view'' of
the evolution of the soliton (in terms of values of $\left| U\right| $), the
horizontal axis being time. The right parts of the panels (sometimes, these
are insets) show a ``top view'' of the evolution, in terms of contour plots
representing the solution. The evolution of solitons in the whole region 
$\kappa \,\,_{\sim }^{<}\,\,0.2$ (note that it comprises both the ``strictly
physical'' and formally unphysical situations) is found to be essentially
the same as shown in Fig. 2 for $\kappa =0.08$.

As it is seen from the plot in Fig.~2(b), and from the results for other
values of $\kappa $, only a single value of $\theta _{\mathrm{in}}$ directly
gives rise to a stable soliton (it is single up to the resolution provided
by the accumulated data, i.e., we cannot rule out that it may be a narrow
interval, rather than the single value). It is also found, more importantly,
that in \emph{all} the cases when the initial pulse relaxes into a stable
soliton, the corresponding final value $\theta _{\mathrm{\ stab}}$ of 
$\theta $ is also a single one, being equal or very close to $\pi /2$, up to
the accuracy of the numerical data. It is noted that $\theta =\pi /2$ yields
the maximum amplitude and minimum width of the soliton as per Eq. (\ref
{pinned}), and it belongs to the stability region (\ref{stability}) of the
unperturbed solitons, although being very close to the instability border.
We also stress that the value $\theta =\pi /2$ always belongs, in the case 
$\lambda =0$, to the existence band (\ref{narrow}) of the pinned-soliton
solutions.

The exact solution (\ref{pinned}) for the trapped soliton with $\theta =\pi
/2 $ takes (provided that $\lambda =0$) a very simple form: as it follows
from Eq. (\ref{a1}), in this case $a=\kappa /2$. Note also that, if $\kappa $
is large, the latter result implies that the maximum value of the soliton's
field is very small: according to Eq. (\ref{centerpoint}), $\left|
U(x=0\right| ^{2}=\left( 4/3\right) \left( \cosh \kappa -1\right) ^{-1}$. In
fact, this means that, in the limit of large $\kappa $, the trapped soliton
becomes a quasi-linear defect-supported mode, cf. Ref. \cite{Weinstein}.

The fact that there is a (presumably) single stable trapped soliton in the
present model implies that the model gives rise to an effective \textit{
attractor}. Although attractors are usually considered as objects specific
to dissipative systems, they may occur in conservative nonlinear-wave models
due to the possibility of radiation losses.

If $\theta _{\mathrm{in}}$ is essentially 
smaller than $\theta _{\mathrm{stab}}$, 
the soliton cannot relax to the stable state and may only decay into
radiation, see Fig.~2(a). In fact, the expression (\ref{E}) for the energy
of the trapped soliton imposes an absolute limit on the values of $\theta_{
\mathrm{in}}^{(0)}$ of the initial unperturbed soliton (the one
corresponding to $\kappa =0$) which can self-trap into the stable state.
Indeed, the energy $E_{\mathrm{in}}^{(0)}$ of the initial unperturbed
soliton is given by Eq. (\ref{E}) with $\kappa =0$, i.e., $E_{\mathrm{in}
}^{(0)}=(8/3)\theta _{\mathrm{in}}^{(0)}$, and, obviously, the energy $E_{
\mathrm{fin}}$ of the established state can only be smaller than $E_{\mathrm{
in}}$. Taking for $E_{\mathrm{fin}}$ the expression (\ref{E}) with $\theta
=\theta _{\mathrm{stab}}\approx \pi /2$, we find $E_{\mathrm{fin}}=(8/3)\sin
^{-1}\left( \mathrm{sech\,}\kappa \right) $. Thus, the limitation $E_{
\mathrm{fin}}<E_{\mathrm{in}}^{(0)}$ takes a final form 
\begin{equation}
\theta _{\mathrm{in}}^{(0)}>\left( \theta _{\mathrm{in}}^{(0)}\right) _{\min
}\equiv \sin ^{-1}\left( \mathrm{sech\,}\kappa \right) \,.  \label{inmin}
\end{equation}
Obviously, $\left( \theta _{\mathrm{in}}^{(0)}\right) _{\min }$
monotonically decreases with $\kappa $, from $\pi /2$ at $\kappa =0$ to 
$0.2244\pi $ when $\kappa =1$ and further to zero as $\kappa \rightarrow
\infty $.

In order to investigate a possible dependence of $\theta _{\mathrm{stab}}$
on the defect's strength $\kappa $, simulations were run for a range of
values of $\theta $ slightly smaller than $\pi /2$, with $\kappa $ varying
from $0.1\pi $ to $0.9\pi $. Similar to what is depicted in Fig.~2(a) and
described above, solitons with $\theta _{\mathrm{in}}<\pi /2$ decay, but at
a slower and slower rate, as $\theta _{\mathrm{in}}$ is getting closer to 
$\pi /2$. For the simulation time $t=300\pi $ and $\kappa =0.1$, the decrease
of the amplitude of $|u|$ and $|v|$ is larger than $1\%$ if $\theta _{
\mathrm{in}}<0.49\pi $. Of course, the decay stops as $\kappa 
\rightarrow0$. However, the decay also becomes slower for 
larger values of $\kappa $
(which can be understood too: in the limit of $\kappa \rightarrow +\infty$,
the trapped soliton goes over into the above-mentioned linear defect mode,
which is stable). For instance, at $\kappa =0.9$, the loss of the amplitude
observed at $t=300\pi $ is less that $1\%$ if $0.47\pi <
\theta _{\mathrm{in}}<\pi /2$. So, there may be two possibilities: 
either $\theta _{\mathrm{stab}
}$ is getting slightly smaller than $\pi /2$ when $\kappa $ is getting
larger, or $\theta _{\mathrm{stab}}$ remains equal to $\pi /2$, but the
solitons are less unstable, so the decay rate for 
$\theta _{\mathrm{in}}<\theta _{\mathrm{stab}}$ is smaller for larger 
values of $\kappa $.
Because of the very slow decay or settling rate of $|u|$ and $|v|$ for 
$\theta _{\mathrm{in}}$ close to $\pi /2$, we cannot pinpoint a single value
of $\theta _{\mathrm{\ stab}}$ value; however, a very conservative estimate
is that, at all the values of $\kappa $ considered, 
$\theta _{\mathrm{\ stab}}$ is confined to the interval $0.47\pi 
<\theta _{\mathrm{\ stab}}<0.53\pi $.

Figure 2(c) shows that if the initial value $\theta _{\mathrm{in}}$ is
somewhat larger than $\theta _{\mathrm{stab}}$, the initial soliton evolves
so that $\theta $ relaxes to $\theta _{\mathrm{stab}}$, shedding off excess
energy in the form of radiation waves. If $\theta _{\mathrm{in}}$ is still
larger for small values of $\kappa $ (actually, if $\theta _{\mathrm{in}
}\,\,_{\sim }^{>}\,\,0.8\pi $ at $\kappa =0.08$, see Fig.~3), the soliton
does not relax into the stable state, but rather decays into radiation,
sometimes generating a small-amplitude residual free soliton which is flung
away from the point where the defect is located [see Fig.~2(d)]; spontaneous
symmetry breaking, apparent in the latter case, is quite possible due to the
instability. However, for $\kappa \,\,_{\sim }^{>}\,\,0.2$, the decay of the
soliton at large values of $\theta _{\mathrm{\ in}}$ is no longer observed.

Figure 3 summarizes the results in the form of a stability region in the
plane ($\kappa ,\theta _{\mathrm{in}}$), in which the initial soliton
evolves into the stable one. It is quite natural that the stability region
expands with the increase of $\kappa $, as it is easier for a stronger
defect to pull the field and thus capture the soliton. Besides that, Eq. 
(\ref{E}) shows that the energy of the trapped soliton decreases with 
$\kappa$, which also makes it easier to create a soliton from a given 
initial pulse.

Lastly, one may notice that, if the value $\theta _{\mathrm{\min }}$
limiting the existence region of the trapped solitons, see Eqs. (\ref{narrow})
and (\ref{min}), exceeds $\theta _{\mathrm{stab}}\approx \pi /2$, the
stable trapped soliton cannot exist. However, in the case under
consideration, $\lambda =0$ and $\kappa >0$, Eq. (\ref{min}) yields $\theta
_{\mathrm{\min }}=2\tan ^{-1}\left( \tanh (\kappa /2)\right) $. Since $\tanh
(\kappa /2)<1$ for all value of $\kappa $, it is obvious that $\theta _{
\mathrm{\min }}$ is always smaller than $\pi /2$ (for example, $\theta _{
\mathrm{\min }}=0.276\pi $ when $\kappa =1$), hence the condition $\theta _{
\mathrm{stab}}>\theta _{\mathrm{\min }}$ never excludes $\theta _{\mathrm{
stab}}$ from the existence region.

\section{Trapping and reflection of a moving soliton by the localized defect}

The interaction of a moving soliton with the defect is a problem of obvious
interest, both in its own right, and for such applications mentioned above
as sensing by means of solitons and the soliton-based dynamical optical
memory. To simulate the interactions, solutions for free moving solitons
were first generated far from the localized defect in a numerical form by
means of the Newton-Raphson method, as stationary solutions in the moving
reference frame (these solutions are also available in the analytical form 
\cite{Aceves,Demetri}, which shows that the soliton's velocity $c$ may take
any value from $\,-1<c<+1$). Then, full simulations of Eqs. (\ref{utux}) and
(\ref{vtvx}) were run by means of the split-step method combined with the
fast Fourier transform, launching the moving soliton to collide it with the
localized defect (it was again set $\lambda =0$ and $\kappa >0$).

First, the parameter plane ($c,\kappa $) was explored, with $c$ taking
values $0.05,\,0.075,\,0.1,\,0.2,\,\ldots ,0.7$ and $\kappa $ ranging from 
$0.07$ to $0.9$. In all the simulations, the soliton was launched with 
$\theta =\pi /2$, which is equal (or very close) to the single value $\theta
_{\mathrm{stab}}$ which, as it was demonstrated above, is selected by the
localized defect as the single value at which the trapped soliton is stable.
Figure~4 displays several different generic examples of the interaction of
the moving soliton with the defect. In each panel, the lower part shows the
evolution of the field $\left| u(x,t)\right| $ by means of contour plots,
and the upper part shows the waveforms $\left| u(x)\right| $ and $\left|
v(x)\right| $ at the end of the simulation, the location of the defect being
marked by a short vertical line.

If the defect is weak, the soliton passes
through it, decreasing its velocity due to radiation loss. However, a
stronger defect captures the soliton into an oscillatory state, see
Fig.~4(a). The possibility of the capture of the soliton (through the
radiation loss) is in agreement with the predicted attractive character of
the interaction between the soliton and the defect, see Eq.~(\ref{U}).

Simulations demonstrate that, quite naturally, the minimum value of the
defect's strength $\kappa $ which is necessary for the capture of a given
free soliton increases with its velocity. Figure 5 shows a border between
the trapping (B) and transmission (A) regions in the plane ($c,\kappa $),
for fixed $\theta =\pi /2$.

However, the soliton-defect interaction is more complicated. When the defect
strength $\kappa $ is large, and when the soliton is energetic enough
[either $c$ or $\theta $ is large -- region (C) in Fig. 5], its energy
splits into three parts as a result of the collision: one part is trapped by
the defect, and two others are scattered away, in both directions.\ The
energy which is not trapped (either bouncing back or the passing through the
defect) self-traps into a secondary soliton. Three examples of this mode of
the interaction are shown in Figs.~4(c,d,e), for fixed values $c=0.5$ and
for $\theta =0.5\pi $, and increasing values of the defect's strength, 
$\kappa =0.3,\,0.6,\,$and $0.8$. It can be seen that when $\kappa $ becomes
larger, shares of the energy which are trapped and reflected back become
larger. In regions A and B in Fig. 5, some energy is also scattered in the
forward or backward directions; however, in these regions scattered energy
is small, and it completely disperses into radiation, without giving rise to
any secondary soliton.

Another characteristic set of numerical data can be displayed for a fixed
small value of the soliton's velocity, $c=0.075$, while the parameter 
$\theta $ of the moving soliton and the defect's strength $\kappa $ are
varied. In this representation, the simulations demonstrate that the defect
behaves in a qualitatively different way for small and large values of 
$\theta $. Namely, for $\theta \,\,_{\sim }^{<}\,\,0.6\pi $, the soliton
interacts with the defect attractively, in accordance with the prediction of
the perturbation theory, while for $\theta \,\,_{\sim }^{>}\,\,0.6\pi $ the
interaction becomes \emph{repulsive}. The change of the sign of the
interaction between the soliton and defect was inferred from the observation
of outcomes of the collision.

In the case $\theta \,\,_{\sim }^{<}\,\,0.6\pi $, the soliton gets captured
by the defect, which clearly testifies that the interaction is attractive.
The capture is accompanied by emission of a
conspicuous jet of radiation. Passage of the soliton through the attractive
defect is also possible (this happens, for instance, if the initial velocity
of the soliton increases).

In a case when the defect strength is $\kappa =0.5$,
and the soliton is heavier, having $\theta =0.7\pi $, the
soliton bounces back from the defect, which is a characteristic feature of
repulsion. The reversal of the interaction from the attraction to repulsion,
which is observed for the solitons with $\theta \,\,_{\sim }^{>}\,\,0.6\pi$,
is beyond the applicability limits of the simple perturbation theory that
gave rise to the effective interaction potential (\ref{U}). However, it
agrees with the findings of the numerical stability analysis reported in the
previous section, where it has been shown that solitons with large initial
values of $\theta _{\mathrm{in}}$ are unstable if trapped by the defect,
which may be realized as a result of repulsion.

Figure~7 summarizes the results obtained for the interaction of the moving
soliton with the local defect in the present case (fixed velocity). The
border between attraction (A) and repulsion (B) regions is shown. To the
left of the regions (A) and (B), there is a thin strip (C) where the defect
is weak, $\kappa \,\,_{\sim }^{<}\,0.1$, and the soliton is heavy enough, 
$\theta \,\,_{\sim }^{>}\,0.45$. In this case, the soliton is neither
captured nor reflected by the defect, but rather passes through it with
significant radiation losses, see Fig.~6. By extrapolating the border
between the regions (A) and (B), it can be inferred that the region (C) is
made up of two sub-regions: an upper one, where the defect is effectively
repulsive, and a lower one, where the defect is attractive. However, in
either case the ``heavy'' soliton passes through the weak defect.

Lastly, in the case when the defect is very strong, and the soliton is very
heavy, the interaction may combine attractive and repulsive features. In
this case, the energy of the incoming soliton is split into three parts: one
is trapped at the defect, while the rest of the energy is scattered in both
the forward and backward directions. The energies scattered away are large
enough for self-trapping into secondary solitons. This happens in the region
(D) of Fig. 7, and is similar to what happens in the region (C) of Fig. 5,
where the defect was very strong, and the soliton was very fast. As the
latter outcome of the collision\ is very similar to what was shown in
Figs.~4(c,d,e), separate figures for this case are not included.

As well as in Fig.~5, the region (D) in Fig. 7 is distinguished as a
separate one to emphasize that \emph{significant} amounts of energy are
scattered away from the defect. In the other regions, some energy is
scattered too, but its amount is much smaller.

Concerning the results presented in this section, it is relevant to mention
that moving unperturbed solitons also have their intrinsic stability limit,
which, however, very weakly depends on the velocity \cite{Barash}: as $c$
increases from $0$ to $1$, the critical value $\theta _{\mathrm{cr}}$, above
which the solitons are unstable, increases from $1.011\cdot \left( \pi
/2\right) $ [see Eq. (\ref{stability})] to $1.017\cdot \left( \pi /2\right) 
$. This implies that, strictly speaking, the incident soliton is unstable in
all the cases shown in Fig. 6; however, its instability is very weak,
therefore it does not manifest itself in any way before the collision is
completed (for the same reason, formally unstable solitons may seem fairly
stable in an experiment).

Finally, we return to the discussion of the validity of Eqs.~(\ref{approxd})
and~(\ref{A}) as an approximation for the defect when $\kappa A>1$, which
corresponds to the formally unphysical situation. We stated above that such
a situation is an appropriate model for a physical one, with a broader
defect and smaller amplitude $A$, so that the product $\kappa \widetilde{
\delta }(x)$ does not exceed $1$, provided that both the ``physical'' and
``unphysical'' perturbations are subject to the identical normalization, 
$\int_{-\infty }^{+\infty }\widetilde{\delta }(x)dx=1$.

As a possibility to verify this statement explicitly, we took the
approximation (\ref{approxd}) with $N$ $=19$ [in the latter case, the
normalization condition (\ref{A}) yields $A=1$], and performed simulations
of the collision of the soliton with this broad defect in the whole $(\kappa
,\theta )$ parameter plane, keeping the soliton's velocity constant, 
$c=0.075 $. The results were then compared to those displayed in Figure~7,
which had been obtained for the narrow defect (recall $N=2$ in the case of
Fig. 7). It was found that, to a large extent, the two sets of the results
are similar indeed. The agreement is very good for larger values of $\kappa$.
Some disagreement occurs in the region (C), where $\kappa $ is small. In
Fig.~7, the vertical part of the border of the region (C) is located at 
$\kappa \simeq 0.1$, with its tip at $\theta \simeq 0.45$. For the broad
defect with $N=19$, it was found that this border moves out to $\kappa
\simeq 0.25$, and the tip of the region moves to $\theta \simeq 0.55$, so
that the region (C) is wider but shorter. This means that the conclusion
whether the solitons can pass through the defect depends not solely on its
integral strength $\kappa \int_{-\infty }^{+\infty }\widetilde{\delta}(x)dx$,
but also on its amplitude $\kappa A$. However, since this affects a small
part of the $(\kappa ,\theta )$ parameter plane, the above conjecture
concerning the actual equivalence of the narrow (formally unphysical) and
broad approximations for the $\delta $-function is quite acceptable.

\section{Conclusion}

In this work, we have studied in detail the dynamics of the interaction of
gap solitons with a localized defect, which is realized as a local
suppression of the Bragg grating that supports the gap solitons. A family of
exact solutions for trapped solitons in the model with the 
$\delta$~-~functional defect, and an approximate expression for the effective
potential of the soliton-defect interaction, were found in the analytical
form. Direct simulations have demonstrated that this conservative model
gives rise to an effective attractor: up to the accuracy provided by
numerical data, the trapped soliton is stable at a single value of its
intrinsic parameter $\theta $ -- in fact, at $\theta =\pi /2$, that
corresponds to the soliton with the largest amplitude and smallest width.
Depending on the strength $\kappa $ of the defect, initial solitons from a
broad domain in the parametric plane ($\kappa ,\theta $) relax into the
stable state; outside the attraction domain, they decay into radiation
(sometimes, generating a residual free soliton).

Simulations of the interaction between moving solitons and the defect
produce a number of different outcomes. If the soliton's velocity and 
$\theta $ are not too large, the soliton is captured by the defect, in
accordance with the fact the effective potential of the interaction between
them is attractive. When the soliton is ``heavy'', with $\theta \,\,_{\sim
}^{>}\,0.6$, the character of the interaction reverses from attraction to
repulsion. In this case, the soliton bounces back. However, if the defect is
weak, regardless of whether the interaction is attractive or repulsive,
heavy solitons can pass through the defect. If it is strong, and the soliton
is heavy or fast enough, the collision may split the soliton in three parts.
Some energy is trapped by the defect to form a pinned soliton, while large
amounts of energy scattered away may generate secondary solitons
(transmitted and reflected ones) of significant amplitude.

These results suggest various applications. Besides the optical-memory
design and sensing (defect detection), which were mentioned in the
Introduction, the delay of a soliton passing through the repulsive defect
can be used to construct an optical delay line. Besides that, an optical
resonator may be implemented by means of a soliton trapped in a cavity
formed by two localized repulsive defects, with a locally enhanced Bragg
reflection. In such setups, the local defects can be switched on and off in
a controllable way by means of electrostriction.

\section*{Acknowledgement}
B.A. Malomed appreciates hospitality of the Department of Electronic
Engineering at the City University of Hong Kong.

\newpage
\section*{Figure Captions}

Fig. 1. The real (even) and imaginary (odd) parts of the field $U(x)$ for
the trapped soliton, with $\protect\kappa = 0.4$ and $\protect\theta = 0.5 
\protect\pi$. The dashed curves depict the exact analytical solution (\ref
{pinned}) obtained for the ideal $\protect\delta$-function. The solid curves
show the numerical solution, with the $\protect\delta$~-~function approximated
as per Eq. (\ref{approxd}) with $N=2$. The two pairs of
curves are nearly indiscernable. It was also checked that using an
approximation closer to the ideal $\protect\delta$~-~function brings the
numerically found waveform still closer to the exact solution.

Fig. 2. Typical results illustrating the stability and instability of the
trapped solitons in the case $\protect\lambda =0$ and 
$\protect\kappa =0.08$. The left part of each panel shows a 
``side view'' of the evolution of $|U|$,
starting with the initial stationary trapped soliton configuration. Right
parts of the panels (or insets) show the evolution in terms of contour
plots. $T$ is the total simulation time. (a) $\protect
\theta_{\mathrm{in}} = 0.4\protect\pi$ is smaller than $\protect\theta_{
\mathrm{stab}}$. The soliton decays into radiation.
(b) $\protect\theta_{\mathrm{in}} = 0.5\protect\pi \approx \protect
\theta_{\mathrm{\ stab}}$. The field directly self-traps into a stable
soliton. 
(c) $\protect\theta =0.7\protect\pi > \protect\theta_{\mathrm{stab}}$. 
The field evolves into a stable soliton, shedding off excess energy in
the form of radiation. (d) $\protect\theta =0.9\protect\pi$. 
This value is much larger
than $\protect\theta_{\mathrm{stab}}$, and the pulse decays into radiation,
generating a small residual soliton thrown away from the defect.

Fig. 3. The region in the plane ($\protect\kappa ,\protect\theta _{\mathrm{
in}}$) between the two borders gives rise to the stable soliton trapped by the
local defect with $\protect\lambda =0$. In fact, in the region of relatively
large values of $\protect\kappa$, the lower border becomes ``fuzzy": as it
is explained in the text, no systematic deviation from $\protect\theta=
\protect\pi/2$ was found, but a deviation within a margin of $\Delta\protect
\theta = 0.03\protect\pi$ is not ruled out, as the instability of unstable
solitons becomes very slow, impeding to monitor their evolution up to a
definite result. Initial solitons taken outside the stability region decay
(in the region above the upper stability border, a residual small-amplitude
soliton can be found, being flung away from the localized defect).

Fig. 4. Collision between a moving soliton with the fixed value $\protect
\theta = \protect\pi/2$ and the defect. The lower and upper panels show,
respectively, the evolution of the field $|u(x,t)|$, and the waveforms 
$|u(x)|$ (solid line) and $|v(x)|$ (dashed line) at the end of the
simulation. (a) The defect with $\protect\kappa=0.2$ captures the
soliton with the velocity $c=0.2$, which is accompanied by conspicuous 
emission of radiation.
(b) In the case of a stronger defect, with $\protect\kappa
=0.6$, less energy passes through, while larger shares of the energy are
trapped at the defect and reflected back.
(c) The case of the strongest defect, with $\protect\kappa=0.8$. 
In this case, most energy is
reflected back, while the amounts of energy trapped and transmitted through
the defect are smaller.

Fig. 5. Regions in the parametric plane ($c,\protect\kappa $), with fixed 
$\protect\theta =\protect\pi /2$, in which the moving soliton, respectively,
passes the defect, is captured by it, or is split into three parts.

Fig. 6. The collision of the moving soliton
with the defect. The parameters are  $c=0.075$,
$\protect\kappa=0.1,\, \protect\theta=0.9\protect\pi$.
The upper and lower panels have the same meaning as in Fig. 4.
In this case, the
soliton overcomes the repulsive barrier and passes through the defect, with
a considerable loss of energy.

Fig. 7. A summary of results obtained for the interaction of the moving
soliton and local defect, in the case when the soliton's velocity is kept
constant at $c=0.075$. In the region A, the defect is attractive for the
solitons with small $\protect\theta $, and the solitons is captured by it.
In the region B, the defect is repulsive for the solitons with larger 
$\protect\theta $; the soliton bounces back. In the region C, the defect is
weak. The solitons passes through it, regardless of whether the interaction
is attractive or repulsive. In the region D, the defect is strong, and,
simultaneously, the soliton is heavy. The interaction combines repulsive and
attractive features, and the soliton splits into three parts. One part is
trapped, the remaining energy being scattered in forward and
backward directions.

\end{document}